\documentclass[journal,twoside]{IEEEtran}
\ifCLASSINFOpdf
\else
\fi
\usepackage[utf8]{inputenc}
\usepackage{multirow}
\usepackage{graphicx}
\usepackage{epstopdf}
\usepackage{stfloats}
\usepackage{subfigure}
\usepackage{microtype}
\usepackage{setspace}
\usepackage{cite}
\usepackage{color}
\usepackage{amsmath}
\usepackage{array}
\usepackage{pifont}
\usepackage{color}
\usepackage[justification=centering]{caption}
\usepackage{bm}
\usepackage{caption}

\begin{document}
%
\title{Map-based Channel Modeling and Generation for U2V mmWave Communication}
\author{Qiuming~Zhu,~\IEEEmembership{Member,~IEEE;} Kai~Mao; Maozhong~Song; Xiaomin~Chen,~\IEEEmembership{Member,~IEEE;} Boyu~Hua; Weizhi~Zhong,~\IEEEmembership{Member,~IEEE;} Xijuan~Ye

}
\maketitle


\begin{abstract}
Unmanned aerial vehicle (UAV) aided millimeter wave (mmWave) technologies have a promising prospect in the future communication networks. By considering the factors of three-dimensional (3D) scattering space, 3D trajectory, and 3D antenna array, a non-stationary channel model for UAV-to-vehicle (U2V) mmWave communications is proposed. The computation and generation methods of channel parameters including inter-path and intra-path are analyzed in detail. The inter-path parameters are calculated in a deterministic way, while the parameters of intra-path rays are generated in a stochastic way. The statistical properties are obtained by using a Gaussian mixture model (GMM) on the massive ray tracing (RT) data. Then, a modified method of equal areas (MMEA) is developed to generate the random intra-path variables. Meanwhile, to reduce the complexity of RT method, the 3D propagation space is reconstructed based on the user-defined digital map. The simulated and analyzed results show that the proposed model and generation method can reproduce  non-stationary U2V channels in accord with  U2V scenarios. The generated statistical properties are consistent with the theoretical and measured ones as well.
\end{abstract}


%
\IEEEpeerreviewmaketitle

\section{Introduction}
%
%
%
%
\IEEEPARstart{U}{nmanned} aerial vehicles (UAVs) are expected to be an important part for the fifth and beyond fifth generation (5G/B5G) communication networks due to its flexible deployment, low cost, and high mobility  \cite{Bin18_ITJ, Wang19_JSAC}. Moreover, the UAV-aided millimeter wave (mmWave) communication system can offer large capacity and good reliability due to its strong line-of-sight (LoS) connection link and wide bandwidth. However, compared with land mobile communication systems, the UAV-to-vehicle (U2V) communications also bring some challenges, i.e., three-dimensional (3D) scattering space, 3D trajectory, and 3D antenna array \cite{Khawaja19_CST, Zhang19_WC, ZhongW19CC}. In order to efficiently design, optimize, and evaluate related communication technologies, it is vital to accurately model and reproduce the U2V mmWave channels\cite{Ullah20_TCCN, ZhuQM18ChCom}.
\par Channel models for UAV-aided communication systems at the sub-6GHz frequency band have been deeply studied during the past decade \cite{WCX19_Access, ChengX19_Commun, Zhu18_CC, ZhangJ19_Access, GuanK20_ArXiv}, but these models cannot be used at the mmWave band directly. Recently, several UAV mmWave channel models based on the geometry-based stochastic model (GBSM) can be addressed in \cite{Zhu19_IET, Mao20_Sensors, Michailidis20_TVT, WCX20_ITJ}. For example, the authors in \cite{Zhu19_IET, Mao20_Sensors} proposed a 3D GBSM for UAV mmWave channels which included the LoS, ground reflection, and single-bounce components. The authors in \cite{Michailidis20_TVT} proposed a 3D regular-shape GBSM for airborne-to-vehicle mmWave communications and analyzed the statistical properties of space-time-frequency correlation. The authors in \cite{WCX20_ITJ} proposed a 3D GBSM for UAV-to-ground multiple-input multiple-output (MIMO) channels and it supported mmWave and massive MIMO configurations. Note that these stochastic models were easy to use but not well consistent with the user-defined specific scenarios. Certainly, field measurement based channel modeling is an accurate method to guarantee realistic channel characteristics for specific scenarios \cite{YinX19_Trans, YangZ18_EuCAP, Matolak17_Trans, Khawaja17_VTC, Geise18CAMA}. However, very few measurement campaigns have been performed for the mmWave band due to its high cost and difficult to implement \cite{Khawaja17_VTC, Geise18CAMA}.
\par On the other hand, the ray tracing (RT) is a good alternative method to obtain accurate channel properties since the mmWave signal has quasi-optical characteristic. Recently, the RT-based channel modeling method has gained more and more attentions \cite{Cheng20Eucap, Khawaja18GSMM, ZhangY19MAP, Alberto19_WM}. Some RT-based channel models can be addressed in \cite{Cheng20Eucap, Khawaja18GSMM} and some propagation characteristics in different scenarios were analyzed in \cite{ZhangY19MAP, Alberto19_WM}. Moreover, the standardized channel model such as METIS also supports a so-called map-based channel model by applying the RT technique on the pre-defined digital maps \cite{METIS15}. However, it should be mentioned that the RT method applying on the original digital map has high complexity and is extremely time-consuming. Meanwhile, it is impractical to trace and generate all rays due to the complexity of digital map. Thus, a hybrid model for terrestrial mobile channels was proposed by the 3GPP group \cite{3GPP18}, but it cannot be applied on the U2V scenarios with the rotation of UAVs. Moreover, the intra-cluster parameter generation method was based on the measurement results of terrestrial mobile systems. This paper aims to fill these research gaps. The main contributions and innovations are summarized as follows:
\par 1) A non-stationary U2V mmWave channel model considering the new factors of 3D scattering space, 3D trajectories, and 3D antenna array is proposed. On this basis, a hybrid framework of channel generator for the pre-defined map is developed with a balance between the accuracy and complexity.
\par 2) A map-based channel parameter computation and generation method is developed. The inter-path parameters are accurately calculated in a deterministic way based on the geometry relationships of specific digital map. The intra-path parameters are generated in a stochastic way. The statistical distributions are obtained by using a Gaussian mixture model (GMM) on the massive RT data in different scenarios. In order to speed up the RT process and reduce the complexity of parameter computation, a 3D reconstruction method for digital map is also given.
\par 3) Based on the proposed model and generation method, a typical UAV mmWave channel in the urban scenario is reproduced, analyzed, and validated. The statistical properties of generated channel, i.e., power delay profile (PDP), autocorrelation function (ACF), and Doppler power spectrum density (DPSD) are also validated by the theoretical and measured data.
\par The rest paper is organized as follows. In Section II, a 3D non-stationary channel model for U2V mmWave communications is proposed. Section III gives the computation and generation method of channel parameters. The simulation and validation of proposed model and generation method are given in Section IV. Finally, conclusions are drawn in Section V.

\section{U2V mmWave channel generation}
\subsection{U2V mmWave channel model}
Let's consider a typical U2V communication scenario, where the UAV and vehicle are configured as the transmitter (TX) and receiver (RX) with their own velocities as
\begin{equation}
{{\bf{v}}^{{\rm{tx/rx}}}}\left( t \right) = \left\| {{{\bf{v}}^{{\rm{tx/rx}}}}} \right\|\left[ {\begin{array}{*{20}{c}}
{\cos \beta _{\rm{v}}^{{\rm{tx/rx}}}(t)\cos\alpha _{\rm{v}}^{{\rm{tx/rx}}}(t)}\\
{\cos \beta _{\rm{v}}^{{\rm{tx/rx}}}(t)\sin \alpha _{\rm{v}}^{{\rm{tx/rx}}}(t)}\\
{\sin \beta _{\rm{v}}^{{\rm{tx/rx}}}(t)}
\end{array}} \right]
\label{0}
\end{equation}
\noindent where $\left\| {{{\bf{v}}^{{\rm{tx/rx}}}}\left( t \right)} \right\|$ represents the velocity magnitude, $\alpha _{\rm{v}}^{{\rm{tx/rx}}}(t)$ and $\beta _{\rm{v}}^{{\rm{tx/rx}}}(t)$ represent the azimuth and elevation angles of velocity, respectively. The detailed parameters are listed in Table I.
\par According to the principle of radio propagation, the receiving signal can be expressed by the summation of $N$ valid paths and each path including ${M_n}$ rays with tiny delay difference \cite{Zhu20IWCMC}. For the 3D scattering space of U2V channels, the channel impulse response (CIR) should consider the impacts of signal propagation on both azimuth and elevation planes. Thus, the time-variant CIR of small scaling fading channel in this paper is modeled as
\begin{equation}
\hspace{-1.5mm}
\begin{array}{c}
h(t,\tau ,{\bf{\Theta }},{\bf{\Phi }}) = \sum\limits_{n = 1}^N {\sum\limits_{m = 1}^{{M_n}} {\{ \sqrt {{P_{n,m}}\left( t \right)} {{\tilde A}_{n,m}}(t)\delta \left( {t - {\tau _{n,m}}\left( t \right)} \right)} } \\
\ \ \ \ \ \ \cdot \delta \left( {{\bf{\Theta }} - {{\bf{\Theta }}_{n,m}}\left( t \right)} \right)\delta \left( {{\bf{\Phi }} - {{\bf{\Phi }}_{n,m}}\left( t \right)} \right)\}
\end{array}
\label{1}
\end{equation}
\noindent where the subscript ${( \cdot )_{n,m}}$ represents the $m{\rm{th}}$ ray within the $n{\rm{th}}$  path, ${P_{n,m}}(t)$, ${\tau _{n,m}}(t)$, ${\tilde A_{n,m}}\left( t \right)$ are the ray power, ray delay and complex ray coefficient, respectively. ${{\bf{\Theta }}_{n,m}}(t) = (\alpha _{n,m}^{{\rm{tx}}}(t), \beta _{n,m}^{{\rm{tx}}}(t))$ and ${{\bf{\Phi }}_{n,m}}(t) = (\alpha _{n,m}^{{\rm{rx}}}(t), \beta _{n,m}^{{\rm{rx}}}(t))$ are the angle of departure (AoD) and angle of departure (AoA), where $\alpha _{n,m}^{{\rm{tx}}}(t)$, $\beta _{n,m}^{{\rm{tx}}}(t)$ denote the azimuth AoD (AAoD) and elevation AoD (EAoD), and $\alpha _{n,m}^{{\rm{rx}}}(t)$ , $\beta _{n,m}^{{\rm{rx}}}(t)$ denote azimuth AoA (AAoA) and elevation AoA (EAoA).
\par For the 3D trajectory and 3D antenna array of U2V channels, the time-variant phase is much more complicated than terrestrial mobile channels. Two factors, i.e., Doppler effect phase caused by 3D trajectory and coordination rotation phase caused by non-linear movement and 3D-shaped array, are considered in this paper, and  ${\tilde A_{n,m}}\left( t \right)$ can be expressed as
\begin{equation}
\hspace{-1mm}
{\tilde A_{n,m}}\left( t \right)\!\! =\!{\!\rm{exp}}\left( {\int_0^t {{\rm{j}}2{\rm{\pi }}{f_{n,m}}\left( {t'} \right)} {\rm{d}}t' + {\rm{j}}\psi _{n,m}^{\rm{R}}(t) + {\rm{j}}\psi^{\rm{I}} _{n,m}} \right) \!\!
\label{2}
\end{equation}
\noindent where $\psi _{n,m}^{\rm{I}} = \int_{ - \infty }^0 {{\rm{j}}2{\rm{\pi }}{f_{n,m}}\left( {t'} \right)} {\rm{d}}t' + \psi^{\rm{'I}} _{n,m}$ is a random phase with initial random phase $\psi^{\rm{'I}} _{n,m}$ , $\psi _{n,m}^{\rm{R}}(t)$ is the time-variant phase caused by rotation, and ${f_{n,m}}(t)$ is the Doppler frequency. In (3), ${f_{n,m}}(t)$ and $\psi _{n,m}^{\rm{R}}(t)$ can be expressed respectively as
\begin{equation}
{f_{n,m}}(t) = \frac{{{f_0}}}{c}\left( {{{\left( {{{\bf{v}}^{{\rm{tx}}}}\left( t \right)} \right)}^{\rm{T}}}{\bf{r}}_{n,m}^{{\rm{tx}}}\left( t \right){\rm{ + }}{{\left( {{{\bf{v}}^{{\rm{rx}}}}\left( t \right)} \right)}^{\rm{T}}}{\bf{r}}_{n,m}^{{\rm{rx}}}\left( t \right)} \right)
\label{3}
\end{equation}
\begin{equation}
\begin{array}{c}
\psi _{n,m}^{\rm{R}}(t) = \frac{{2\pi {f_0}}}{c}\left( {{\bf{r}}_{n,m}^{{\rm{rx}}}(t) \cdot {{\bf{R}}^{{\rm{rx}}}}(t) \cdot {{\bf{d}}^{{\rm{rx}}}}(t)} \right)\\
\ \ \ \ \ \ \ \ \ \ \ + \frac{{2\pi {f_0}}}{c}\left( {{\bf{r}}_{n,m}^{{\rm{tx}}}(t) \cdot {{\bf{R}}^{{\rm{tx}}}}(t) \cdot {{\bf{d}}^{{\rm{tx}}}}(t)} \right)
\end{array}
\label{4}
\end{equation}
\noindent where $c$ is light speed, ${f_0}$ is the carrier frequency, ${\bf{r}}_{n,m}^{{\rm{tx/rx}}}(t)$, ${{\bf{R}}^{{\rm{tx/rx}}}}(t)$ and ${{\bf{d}}^{{\rm{tx/rx}}}}(t)$ are the normalized direction vector, rotation matrix, and location vector, respectively. In (5), ${\bf{r}}_{n,m}^{{\rm{tx/rx}}}(t)$ and ${{\bf{R}}^{{\rm{tx/rx}}}}(t)$ can be further described as (6) and (7).
\begin{equation}
{\bf{r}}_{n,m}^{{\rm{tx/rx}}}(t) = \left[ {\begin{array}{*{20}{c}}
{\cos \beta _{n,m}^{{\rm{tx/rx}}}(t)\cos\alpha _{n,m}^{{\rm{tx/rx}}}(t)}\\
{\cos \beta _{n,m}^{{\rm{tx/rx}}}(t)\sin \alpha _{n,m}^{{\rm{tx/rx}}}(t)}\\
{\sin \beta _{n,m}^{{\rm{tx/rx}}}(t)}
\end{array}} \right]
\label{5}
\end{equation}

\begin{figure*}[!t]
\normalsize
\begin{equation}
{{\bf{R}}^{{\rm{tx/rx}}}}(t) =\left[ {\begin{array}{*{20}{c}}
{\cos \alpha _{\rm{v}}^{{\rm{tx/rx}}}(t)\cos \beta _{\rm{v}}^{{\rm{tx/rx}}}}&\!\!\!\!\!\!{ - \sin \alpha _{\rm{v}}^{{\rm{tx/rx}}}}&\!\!\!\!{ - \cos \alpha _{\rm{v}}^{{\rm{tx/rx}}}\sin \beta _{\rm{v}}^{{\rm{tx/rx}}}}\\
{\sin \alpha _{\rm{v}}^{{\rm{tx/rx}}}\cos \beta _{\rm{v}}^{{\rm{tx/rx}}}}&\!\!\!{\cos \alpha _{\rm{v}}^{{\rm{tx/rx}}}}&\!\!\!\!{ - \sin \alpha _{\rm{v}}^{{\rm{tx/rx}}}\sin \beta _{\rm{v}}^{{\rm{tx/rx}}}}\\
{\sin \beta _{\rm{v}}^{{\rm{tx/rx}}}}&\!\!\!0&\!\!\!\!{\cos \beta _{\rm{v}}^{{\rm{tx/rx}}}}
\end{array}} \right]
\label{6}
\end{equation}
\end{figure*}

\begin{table}[!b]
\centering
\caption{Parameter definitions in the proposed model}
\centering
\renewcommand\arraystretch{1.3}
\begin{tabular}{p{3cm}<{\centering}p{5.5cm}<{\centering}}
\hline
\ Parameters & Definition  \\ \hline
\ $N$ & path number \\
\ ${M_n}$ & intra-path ray number\\
\ ${P_{n,m}}(t),{\tau _{n,m}}(t)$ & ray power and delay, respectively \\
\ ${M_n}$ & path number \\
\ ${\tilde A_{n,m}}\left( t \right)$ & ray complex coefficient \\
\ ${\bf{\Theta}}_{n,m}(t),{\bf{\Phi}}_{n,m}(t)$ & AoD and AoA, respectively \\
\ ${f_{n,m}}(t)$ & Doppler frequency \\
\ $\psi _{n,m}^{\rm{I}}$ & initial phase \\
\ $\psi _{n,m}^{\rm{R}}(t)$ & phase caused by antenna array rotation \\
\ ${\bf{r}}_{n,m}^{{\rm{tx/rx}}}(t)$ & normalized direction vector  \\
\ ${{\bf{d}}^{{\rm{tx/rx}}}}(t)$ & location vector of TX (or RX) \\
\ ${{\bf{R}}^{{\rm{tx/rx}}}}(t)$ & rotation matrix of TX (or RX) \\
\hline
\end{tabular}
\label{table I}
\end{table}

\subsection{ Framework of channel generator}
The proposed channel model has unique inter-path and intra-path characteristics due to high frequency with large bandwidth and high time resolution. The procedure of channel generation combining with RT method is shown in Fig. 1. Firstly, the inter-path parameters can be calculated based on the path tracing according to the geometric relationships. Secondly, massive data of intra-path ray parameters is obtained and used to extract the statistical distributions, i.e., delay offsets and angle offsets under different type scenarios. On this basis, the random ray parameters can be generated during each simulation. Finally, the U2V channels can be reproduced by combing the channel parameters with proposed model.
\par Considering that the realistic scenario contains abundant scatterers, the ray tracing process for the original map is extraordinarily complex and time-consuming. Thus, a reconstruction method is developed in this paper. Firstly, the digital terrain model (DEM) of user-defined scenario can be obtained from geographic database such as Google earth. The DEM file contains the information of latitude, longitude, and elevation of all points. These information is then used to reconstruct many regular or irregular triangle facets to approximately reconstruct the surface of 3D terrain and buildings.

\begin{figure}[!b]
\setlength{\abovecaptionskip}{1.3cm}
\setlength{\belowcaptionskip}{0.3cm}
	\centering
	\includegraphics[width=85mm]{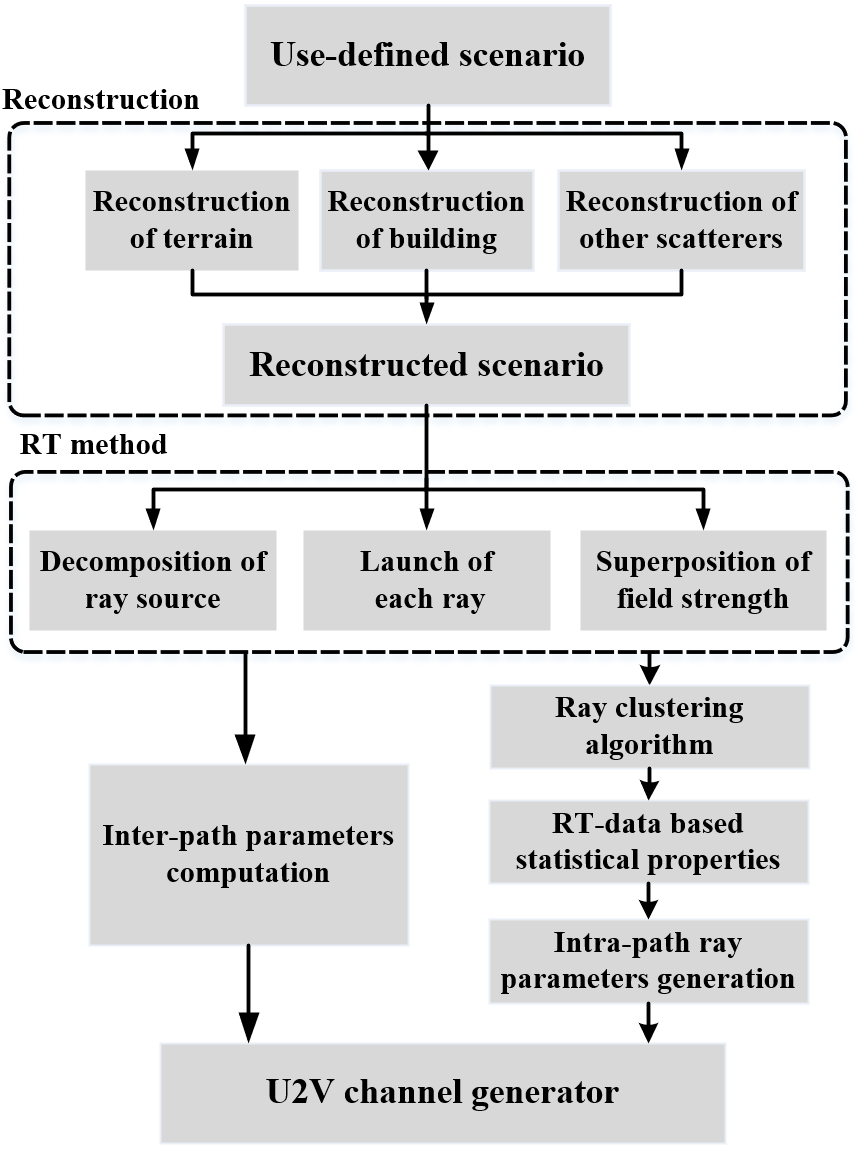}
	\caption{ Procedure of channel generation.}
    \label{fig.1}
\end{figure}

\section{Channel parameter computation and generation}
\subsection{RT-based path parameter computation}
The mmWave signal propagation process has apparent quasi-optical characteristic. Based on the ray optics and uniform theory of diffraction theory, the RT method can precisely analyze the mmWave propagation process with a specific map. The proposed RT-based parameter computation method includes three steps, i.e., decomposition of ray source, tracking of rays, and superposition of filed strength. Since it is impossible to track all the rays from the signal source, we choose the rays uniformly distributed on the sphere. The azimuth and elevation angles of candidate rays can be expressed respectively as
\begin{equation}
\begin{array}{l}
\varphi  = i\delta ,i = 1 \cdots {\mathop{\rm int}} (180/\delta )\\
\theta  = j\delta /\sin (\varphi ),j = 1 \cdots {\mathop{\rm int}} (360\sin (\varphi )/\delta )
\end{array}
\label{7}
\end{equation}
\noindent where $\delta $ denotes the angle space, int$\left(  \cdot  \right)$  means the integer part. In order to track the full propagation trace of candidate rays, it's important to find the reflected point that the rays hit. Since each reconstructed scatterer consists of several triangle facet, we denote the target triangle facet as ${V_0}{V_1}{V_2}$ and the location vector of ${V_0}$, ${V_1}$, ${V_2}$ are denoted as ${{\bf{V}}_0}$, ${{\bf{V}}_1}$ and ${{\bf{V}}_2}$ , respectively. The location vector ${\bf{P}}$ of arbitrary point $P$ in triangle facet can be expressed as
\begin{equation}
{\bf{P}} = {p_0}{{\bf{V}}_0} + {p_1}{{\bf{V}}_1} + {p_2}{{\bf{V}}_2}
\label{8}
\end{equation}
\noindent where ${p_0}$, ${p_1}$, ${p_2}$ can be further determined by
\begin{equation}
\left\{ {\begin{array}{*{20}{c}}
{{p_0} + {p_1} + {p_2} = 1}\\
{{p_0} = \frac{{S\left( {P{V_1}{V_2}} \right)}}{{S\left( {{V_0}{V_1}{V_2}} \right)}},{p_0} \in [0,1]}\\
{{p_1} = \frac{{S\left( {P{V_0}{V_2}} \right)}}{{S\left( {{V_0}{V_1}{V_2}} \right)}},{p_1} \in [0,1]}\\
{{p_2} = \frac{{S\left( {P{V_0}{V_1}} \right)}}{{S\left( {{V_0}{V_1}{V_2}} \right)}},{p_2} \in [0,1]}
\end{array}} \right.
\label{9}
\end{equation}
\noindent where $S\left(  \cdot  \right)$ is the triangle area. Let us denote the location of signal source as ${\bf{Q}}$, then the location of each point along the ith ray can be expressed as
\begin{equation}
{{\bf{I}}_i} = {\bf{Q}} + {w_1}{{\bf{S}}_i}
\label{10}
\end{equation}
\noindent where ${w_1}$ is the length, ${{\bf{S}}_i}$ is the spherical unit vector and can be expressed as
\begin{equation}
{{\bf{S}}_i} = {\left[ {\begin{array}{*{20}{c}}
{\sin\varphi \cos\theta }&{\sin \varphi \sin\theta }&{ - \cos \varphi }
\end{array}} \right]^{\rm{T}}.}
\label{11}
\end{equation}
\par By using the Snell's law, the unit vector of ith reflected ray can be expressed as ${{\bf{P}}_i}$ is a possible reflected point, it can be determined by
\begin{equation}
{\bf{S}}_i^{{\mathop{\rm Ref}\nolimits} } = {{\bf{S}}_i} - 2({{\bf{S}}_i} \cdot {\bf{k}}){\bf{k}}
\label{12}
\end{equation}
\noindent where ${\bf{k}}$ is the unit vector of triangle facet. Assuming that ${{\bf{P}}_i}$ is a possible reflected point, it can be determined by
\begin{equation}
\left\{ \begin{array}{l}
{{\bf{P}}_i} = {{\bf{I}}_i}\\
{{\bf{P}}_i} = {\bf{R}} - {\left\| {{{\bf{P}}_i} - {\bf{R}}} \right\|_2}{\bf{S}}_i^{{\mathop{\rm Ref}\nolimits} }
\end{array} \right.,{{\bf{P}}_i} \in {\bf{P}}
\label{13}
\end{equation}
\noindent where ${w_2}$ is the ray length from the reflected point to the RX, and ${\bf{R}}$ is the location vector of RX.
\par The arrival rays at the RX usually include two cases, e.g., direction and reflection. For the direct case, the electric field intensity can be calculated by
\begin{equation}
{\bf{E}}_{}^{{\rm{LoS}}}(t) = {{\bf{E}}^{1{\rm{m}}}}\frac{{{{\rm{e}}^{ - {\rm{j}}ld_0^{{\rm{tx}}}(t)}}}}{{d_0^{{\rm{tx}}}(t)}}
\label{14}
\end{equation}
\noindent where ${\bf{E}^{{\rm{1m}}}}$ is the electric field intensity of 1 m from the TX, $l = 1/{\lambda _0}$ is the wave number, and $d_0^{{\rm{tx}}}(t) = {\left\| {{\bf{Q}} - {\bf{R}}} \right\|_2}$ is the distance between the UAV and RX. For the reflected case, it can be obtained by
\begin{equation}
{\bf{E}}_m^{\rm{R}}(t) = {{\bf{E}}^{1{\rm{m}}}}R\frac{{{{\rm{e}}^{ - lj(d_m^{{\rm{tx,S}}}(t) + d_m^{{\rm{S,rx}}}(t))}}}}{{d_m^{{\rm{tx,S}}}(t) + d_m^{{\rm{S,rx}}}(t)}}
\label{15}
\end{equation}
\noindent where $R$ is the reflection coefficient, $d_m^{{\rm{tx,S}}}(t) = {\left\| {{\bf{Q}} - {{\bf{P}}_i}} \right\|_2}$ is the distance between the UAV and the scatterer, and $d_m^{{\rm{S,rx}}}(t) = {\left\| {{{\bf{P}}_i} - {\bf{R}}} \right\|_2}$ is the distance between the RX and scatterer. Finally, the power gain of mth ray can be obtained by
\begin{equation}
{P_m} = 10{\log _{10}}\left( {{G^{{\rm{rx}}}}{G^{{\rm{tx}}}}{{(\frac{{{\lambda _0}}}{{4\pi }})}^2}\left| {\frac{{{\bf{E}}_m^{\rm{R}}}}{{{{\bf{E}}_{1{\rm{m}}}}}}} \right|} \right)
\label{16}
\end{equation}
\noindent where ${G^{{\rm{tx}}}}$ and ${G^{{\rm{rx}}}}$ are antenna gains of TX and RX, respectively. The power gain of direction ray can be calculated similarly.

\par Note that each non-light-of-sight (NLoS) path contains several intra-path rays and the arrival rays from the same scatterer have similar delays. In this paper, the ray delays within each path are modeled by adding an intra-path delay offset $\Delta {\tau _{n,m}}$ on the mean path delay $\bar \tau _n^{}\left( t \right)$ as
\begin{equation}
\tau _{n,m}^{}\left( t \right) = \bar \tau _n^{}\left( t \right) + \Delta {\tau _{n,m}}.
\label{17}
\end{equation}
\noindent Similarly, the ray angles $\alpha _{n,m}^{{\rm{rx}}}\left( t \right)$, $\beta _{n,m}^{{\rm{rx}}}\left( t \right)$ can be modeled as several intra-path offsets $\Delta {\alpha _{n,m}}$ , $\Delta {\beta _{n,m}}$ deviating from the mean path angle $\bar \alpha _n^{{\rm{rx}}}\left( t \right)$, $\bar \beta _n^{{\rm{rx}}}\left( t \right)$, which have similar models as (18).
\par In this paper, the inter-path delays and angles are calculated in a deterministic way. The adjacent reflection points are assumed to be one centroid and the location vector is denoted as ${{\bf{\bar P}}_n}$. The mean values of path delay, AAoA, and EAoA can be expressed respectively as (19),(20) and (21).
\begin{equation}
\bar \tau _n^{}\left( t \right) = \frac{{{{\left\| {{\bf{Q}} - {{{\bf{\bar P}}}_n}} \right\|}_2} + {{\left\| {{{{\bf{\bar P}}}_n} - {\bf{R}}} \right\|}_2}}}{c}
\label{18}
\end{equation}
\begin{figure*}[!t]
\normalsize
\begin{equation}
\bar \alpha _n^{{\rm{rx}}}\left( t \right) = \left\{ {\begin{array}{*{20}{c}}
\!\!{\arccos (\frac{{\left| {{{{\bf{\bar P}}}_n}^x - {{\bf{R}}^x}} \right|}}{{\sqrt {{{\left| {{{{\bf{\bar P}}}_n}^x - {{\bf{R}}^x}} \right|}^2} + {{\left| {{{{\bf{\bar P}}}_n}^y - {{\bf{R}}^y}} \right|}^2}} }}),{{{\bf{\bar P}}}_n}^x - {{\bf{R}}^x} \ge 0}\\
\!\!{\pi  - \arccos (\frac{{\left| {{{{\bf{\bar P}}}_n}^x - {{\bf{R}}^x}} \right|}}{{\sqrt {{{\left| {{{{\bf{\bar P}}}_n}^x - {{\bf{R}}^x}} \right|}^2} + {{\left| {{{{\bf{\bar P}}}_n}^y - {{\bf{R}}^y}} \right|}^2}} }}),{{{\bf{\bar P}}}_n}^x - {{\bf{R}}^x} < 0}
\end{array}} \right.
\label{19}
\end{equation}
\end{figure*}
\begin{equation}
\bar \beta _n^{{\rm{rx}}}\left( t \right) = \arcsin \left( {\frac{{\left| {{{{\bf{\bar P}}}_n}^z - {{\bf{R}}^z}} \right|}}{{{{\left\| {{{{\bf{\bar P}}}_n} - {\bf{R}}} \right\|}_2}}}} \right)
\label{20}
\end{equation}
\noindent where ${\left(  \cdot  \right)^x}$ , ${\left(  \cdot  \right)^y}$ and ${\left(  \cdot  \right)^z}$ represent the x, y, and z component, respectively. Similarly, the mean path power ${\bar P_n}(t)$ can be calculated by (15)$-$(17).
\subsection{ RT-statistical parameters and generation}
There is no need to calculate all ray parameters in each channel modeling or generation for three reasons. Firstly, each scenario includes numerous scatterer elements, e.g., vegetation, puddles, roads, and different roof structures, which may result in several hours of computation time. Secondly, it is impossible to describe the realistic propagation scenario perfectly. For example, the material of composite scatterers, irregular building surface, branches and leaves of trees are difficult to describe. Thirdly, there are lots of scatterers such as trains, vehicles, and pedestrian that are usually in motion and do not have fixed positions. In the proposed generator, the ray parameters are generated in a stochastic way based on the empirical statistical properties.
\par In order to obtain the statistical properties of intra-path ray parameters, four typical realistic scenarios included in most of standardized channel models, i.e., urban, hilly, forest, and sea, are studied in this section. We set 50 ground RXs at the height of 2 m and 500 UAV-based TXs at the height of 150 m in each scenario. By applying the RT method, the U2V channels between 25000 pairs of transceivers are obtained in different scenarios, where each channel contains massive propagation rays. Furthermore, the ray parameters, i.e., delays, angles and powers, are analyzed to get the statistical properties of intra-path ray offsets. The acquisition RT-based data under urban scenario is shown in Fig. 2.
\begin{figure}[!b]
	\centering
	\includegraphics[width=85mm]{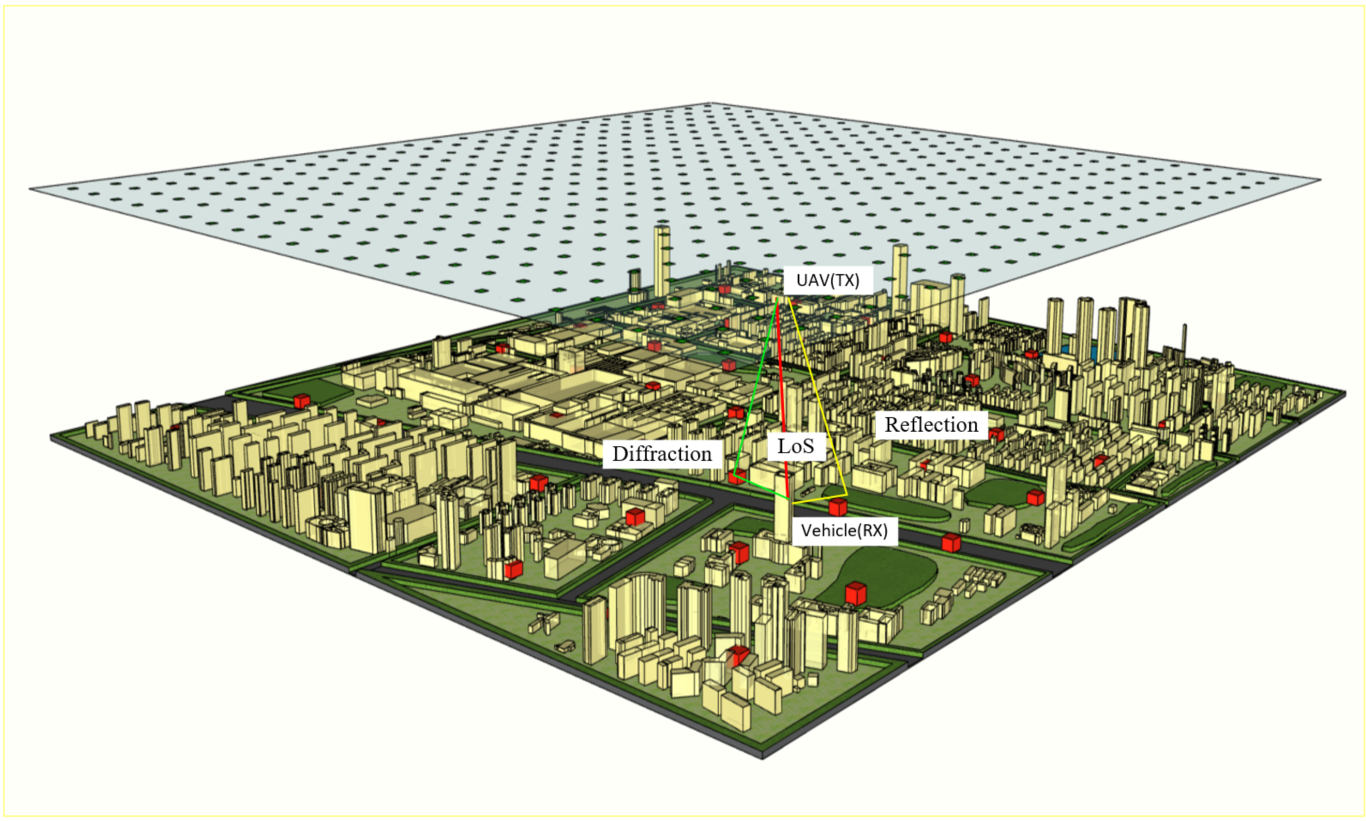}
	\caption{Acquisition of massive RT-based data.}
    \label{fig:2}
\end{figure}
\par To further analyze the intra-path ray parameters, the key step is to group the rays into several clusters with similar characteristics, e.g., delay and AoA. Here, the distance between two rays is defined by the multiple component distance (MCD) as (22).
\begin{figure*}[!t]
\normalsize
\begin{equation}
{\rm{MCD(}}{m_i},{m_j}{\rm{)}} = \sqrt {{{\left\| {{\rm{MC}}{{\rm{D}}_\tau }{\rm{(}}{m_i},{m_j}{\rm{)}}} \right\|}^2} + {{\left\| {{\rm{MC}}{{\rm{D}}_{{\rm{AoA}}}}{\rm{(}}{m_i},{m_j}{\rm{)}}} \right\|}^2}} {\rm{ }}
\label{21}
\end{equation}
\end{figure*}

\par On this basis, ${\rm{MC}}{{\rm{D}}_\tau }{\rm{(}}{m_i},{m_j}{\rm{)}}$ and ${\rm{MC}}{{\rm{D}}_{{\rm{AoA}}}}{\rm{(}}{m_i},{m_j}{\rm{)}}$ can be obtained respectively by (23) and (24).
\begin{equation}
{\rm{MC}}{{\rm{D}}_\tau }{\rm{(}}{m_i},{m_j}{\rm{)}} = \zeta \frac{{\left| {{\tau _{{m_i}}} - {\tau _{{m_j}}}} \right|{\tau _{std}}}}{{\Delta \tau _{\max }^2}}
\label{22}
\end{equation}
\begin{figure*}[!t]
\normalsize
\begin{equation}
{\rm{MC}}{{\rm{D}}_{{\rm{AoA}}}}{\rm{(}}{m_i},{m_j}{\rm{)}} =\! \frac{1}{2}{\left\| {\left[ {\begin{array}{*{20}{c}}
\!{\sin{\beta _{{m_i}}}\cos{\alpha _{{m_i}}}}\\
\!{\sin {\beta _{{m_i}}}\sin{\alpha _{{m_i}}}}\\
\!{ - \cos {\beta _{{m_i}}}}
\end{array}} \right]\!\! -\!\! \left[ {\begin{array}{*{20}{c}}
\!{\sin{\beta _{{m_j}}}\cos{\alpha _{{m_j}}}}\\
\!{\sin {\beta _{{m_j}}}\sin{\alpha _{{m_j}}}}\\
\!{ - \cos {\beta _{{m_j}}}}
\end{array}} \right]} \right\|_2}
\label{23}
\end{equation}
\end{figure*}\\
where $\zeta $ is the delay factor, ${\tau _{{m_i}/{m_j}}}$ is the ray delay, ${\tau _{std}}$ is the standard deviation of ray delays, and $\Delta \tau _{\max }^{}$ is the maximum value of relative delays. In (24), ${\alpha _{{m_i}/{m_j}}}$ and ${\beta _{{m_i}/{m_j}}}$ represent the AAoA and EAoA, respectively. When the MCD of two rays is smaller than the threshold ${\rm{MC}}{{\rm{D}}_{thr}}$ , they would be grouped into the same cluster. Algorithm 1 gives the procedure of our new clustering method, where ${{\bf{{\rm X}}}_m} = \left\{ {{P_m},{\tau _m},{\alpha _m},{\beta _m}\left| {m = 1,2,...,M} \right.} \right\}$ is the aggregate containing all the rays and ${{\bf{{\rm X}}}_n}$ is the aggregate containing the rays within the $n$th path.
\begin{table}[h]
\centering
\centering
\renewcommand\arraystretch{1.2}
\begin{tabular}{p{8cm}}
\hline
$\textbf{Algorithm 1}$:  Ray clustering method \\ \hline
$\textbf{Input}$: Aggregate ${{\bf{{\rm X}}}_m}$ ; \\
$\textbf{Output}$: Aggregate ${{\bf{{\rm X}}}_n}$ ;\\
\ 1:$\textbf{Initialize}$   N = 0, n = 1, $\zeta $ = 3,${\tau _{std}} = {\rm{std}}\left( {{\tau _m}} \right)$ ; \\
\ 2:\ \ $\Delta \tau _{\max }^{} = \max \{ \left| {{\tau _{{m_i}}} - {\tau _{{m_j}}}} \right|\} $, ${\tau _{{m_i}}}$, ${\tau _{{m_j}}}$$ \in {{\bf{X}}_m}$ ;\\
\ 3:$\textbf{while}$ ${{\bf{{\rm X}}}_m} \not\subset \emptyset $ $\textbf{do}$\\
\ 4:\ \ \ \ k = find(${P_m}$ == max(${P_m}$)),${P_m} \in {{\bf{X}}_m}$; \\
\ 5:\ \ \ \ $\textbf{for}$ m = 1 : card(${{\bf{{\rm X}}}_m}$) $\textbf{do}$ \\
\ 6:\ \ \ \ \ \ $\textbf{if}$ ${\rm{MCD(}}m,k{\rm{)}} \le {\rm{MC}}{{\rm{D}}_{thr}}$ $\textbf{then}$ \\
\ 7:\ \ \ \ \ \ \ \ ${{\bf{{\rm X}}}_m} = {{\bf{{\rm X}}}_m} - \left\{ {{P_m},{\tau _m},{\alpha _m},{\beta _m}} \right\}$; \\
\ 8:\ \ \ \ \ \ \ \ ${{\bf{{\rm X}}}_n} = {{\bf{{\rm X}}}_n} + \left\{ {{P_m},{\tau _m},{\alpha _m},{\beta _m}} \right\}$ ;\\
\ 9:\ \ \ \ \ \ $\textbf{end if}$\\
\ 10:\ \ \ \ $\textbf{end for}$ \\
\ 11:\ \ \ \ $n = n + 1$;\\
\ 12: $\textbf{end while}$\\
\hline
\end{tabular}
\label{table I}
\vspace{-0.8cm}  
\setlength{\abovecaptionskip}{-0.2cm}   
\setlength{\belowcaptionskip}{-0.8cm}   
\end{table}
\par By applying the ray clustering method on massive channel data, huge amount of output ${{\bf{{\rm X}}}_n}$ is analyzed to get the statistical properties of ray parameters under different scenarios. The delay offset of rays deviating from the mean path delay are shown in Fig. 3. As a general and classic distribution, the Gaussian distribution (GD) is previously used to fit the distribution of channel parameters \cite{3GPP18}. However, we have found that GD cannot fit well with the PDF of ray delay offset. Instead, a Gaussian mixture model (GMM) is adopted in this paper. As we can see in Fig. 3, the GMM fits better than the traditional GD. To further evaluate the goodness of fit, the root mean square errors (RSMEs) of GMM and GD are calculated by
\begin{equation}
{R_e} = \!\! \sqrt{\frac{1}{K}\sum\limits_{k = 1}^K {{{\left( {{y_k}\!\! -\!\! {{\hat y}_k}} \right)}^2}}}\!\!
\label{23}
\end{equation}
\noindent where ${y_k}$, ${\hat y_k}$  represent the PDF of RT data and fitted data, respectively. InIn Fig. 3, ${R_e}$ of GMM are 0.0021, 0.0052, 0.009, 0.001 under four scenarios, which outperform the 0.0317, 0.0341, 0.0195, 0.021 of GD. For this reason, the ray delay offset is modeled as
\begin{equation}
{f_{\Delta \tau }}(\tau ) = \sum\limits_{i = 1}^I {\frac{{{a_i}}}{{\sqrt {2\pi } {\sigma _i}}}\exp ( - \frac{{{{(\tau  - {\mu _i})}^2}}}{{2{\sigma _i}^2}})}
\label{25}
\end{equation}
\noindent where $I$ is 2 in this paper, ${a_i}$ is the weight factor of the magnitude, ${\mu _i}$ is the mean value and $\sigma _i^{}$ is the standard deviation. The fitting parameters of ${a_1}$, ${\mu _1}$, $\sigma _1^{}$, ${a_2}$, ${\mu _2}$ and $\sigma _2^{}$ are listed in Table II.
\begin{figure}[!t]
\setlength{\belowcaptionskip}{0cm}
	\centering
	\includegraphics[width=85mm]{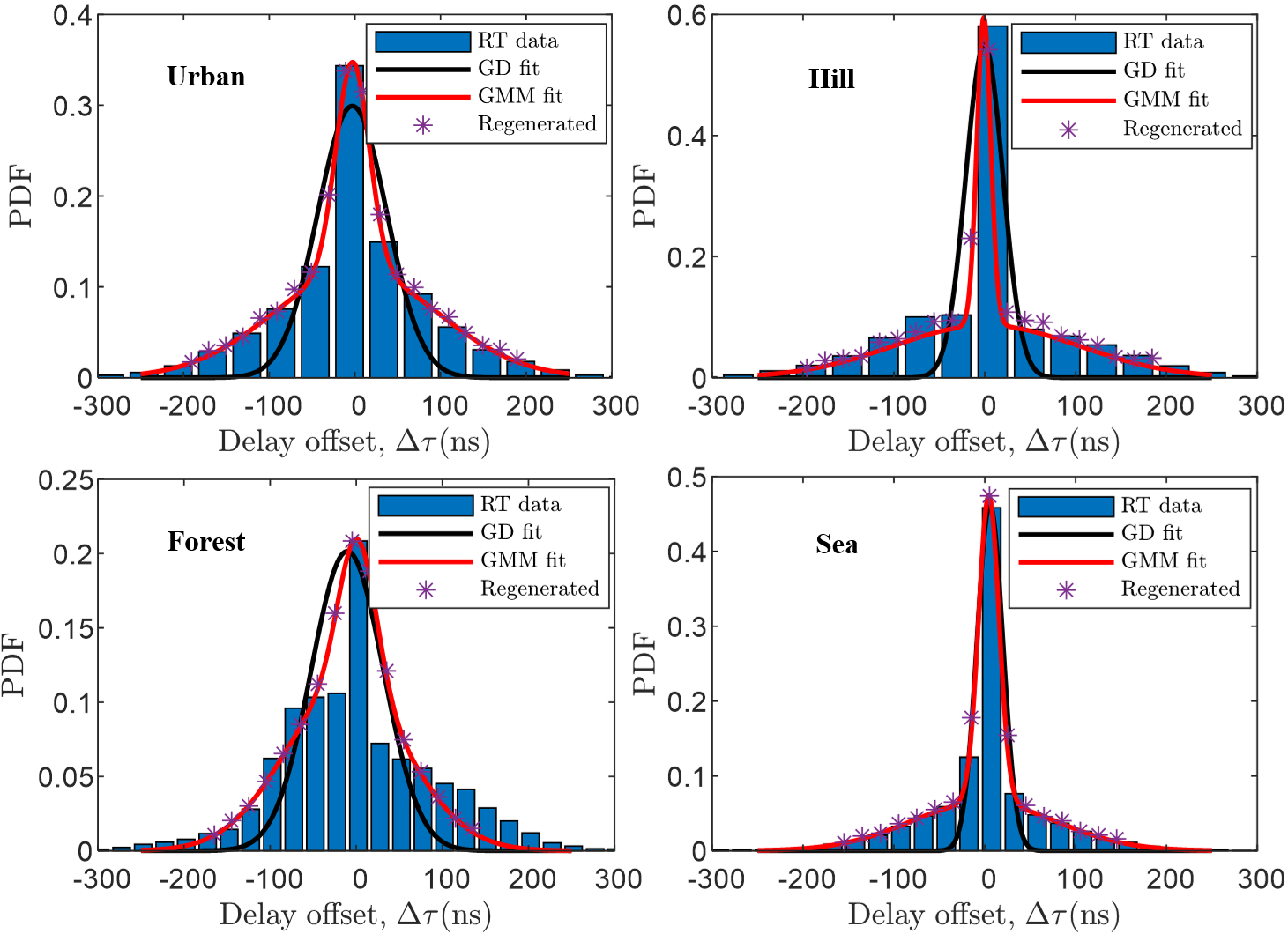}
	\caption{PDFs of ray delay offset.}
    \label{fig:3}
\end{figure}
\par On the other hand, the Von Mise distribution (VMD) is widely used to descript the distribution of AoA and AoD \cite{Ma_20}. However, we also find that the PDFs of azimuth and elevation angle offset can be fitted better by the GMM as shown in Fig. 4 and Fig. 5. In Fig. 4, ${R_e}$ of GMM are 0.0028, 0.0009, 0.0024, 0.0022 under four scenarios, which outperform the 0.0301, 0.0505, 0.0274, 0.0401 of VMD. In Fig. 5, ${R_e}$ of GMM are 0.0023, 0.0018, 0.0014, 0.0013, which outperform the 0.0115, 0.0170, 0.0220, 0.0163 of VMD. The fitting parameters of the PDFs of azimuth and elevation angle offset are also listed in Table II.
\begin{table}[!t]
\centering
\caption{Fitting parameters of PDFs}
\centering
\begin{tabular}{|p{1.3cm}<{\centering}|p{0.3cm}<{\centering}|p{1.1cm}<{\centering}|p{1.1cm}<{\centering}|p{1.1cm}<{\centering}|p{1.1cm}<{\centering}|}
\hline
\multicolumn{2}{|c|}{Scenarios} & Urban & Hilly & Forest & Sea  \\ \hline
\multirow{6}{1cm}{Delay offset} & ${a_1}$ & 11.244 & 9.751 & 5.577 & 11.522 \\ \cline{2-6}
 & ${\mu _1}$ & -2.747 & -1.090 & 1.391 & 4.438 \\ \cline{2-6}
 & $\sigma _1^{}$ & 19.410 & 7.644 & 21.748 & 11.201\\ \cline{2-6}
 & ${a_2}$ & 56.199 & 22.170 & 19.373 & 12.998\\ \cline{2-6}
 & ${\mu _2}$ & 0.903 & 1.761 & -11.780 & -1.332\\ \cline{2-6}
 & $\sigma _2^{}$ & 97.015 & 102.248 & 70.711 & 80.186\\ \cline{2-6}\hline
\multirow{6}{1cm}{Azimuth offset} & ${a_1}$ & 0.519 & 0.584 & 0.379 & 0.432\\ \cline{2-6}
 &  ${\mu _1}$ & 0.013 & 0.630 & 0.191 & 0.201 \\ \cline{2-6}
 &  $\sigma _1^{}$ & 0.886 & 0.712 & 0.588 & 0.404 \\ \cline{2-6}
 &  ${a_2}$ & 1.259 & 1.431 & 0.809 & 0.394 \\ \cline{2-6}
 &  ${\mu _2}$ & -0.083 & -0.054 & -0.015 & -0.004 \\ \cline{2-6}
 &  $\sigma _2^{}$ & 4.431 & 5.685 & 3.471 & 3.386 \\ \cline{2-6}\hline
\multirow{6}{1cm}{Elevation offset} & ${a_1}$ & 0.371 & 0.984 & 0.642 & 0.629\\ \cline{2-6}
 &  ${\mu _1}$ & -0.131 & 0.244 & 0.167 & 0.069 \\ \cline{2-6}
 &  $\sigma _1^{}$ & 0.565 & 0.699 & 0.437 & 0.354 \\ \cline{2-6}
 &  ${a_2}$ & 0.476 & 0.345 & 0.162 & 0.138 \\ \cline{2-6}
 &  ${\mu _2}$ & 0.027 & -1.345 & -1.323 & -0.107 \\ \cline{2-6}
 &  $\sigma _2^{}$ & 2.210 & 4.295 & 5.046 & 2.303 \\ \cline{2-6}
\hline
\end{tabular}
\label{table II}
\end{table}
\begin{figure}[!t]
\setlength{\belowcaptionskip}{-0.3cm}
	\centering
	\includegraphics[width=85mm]{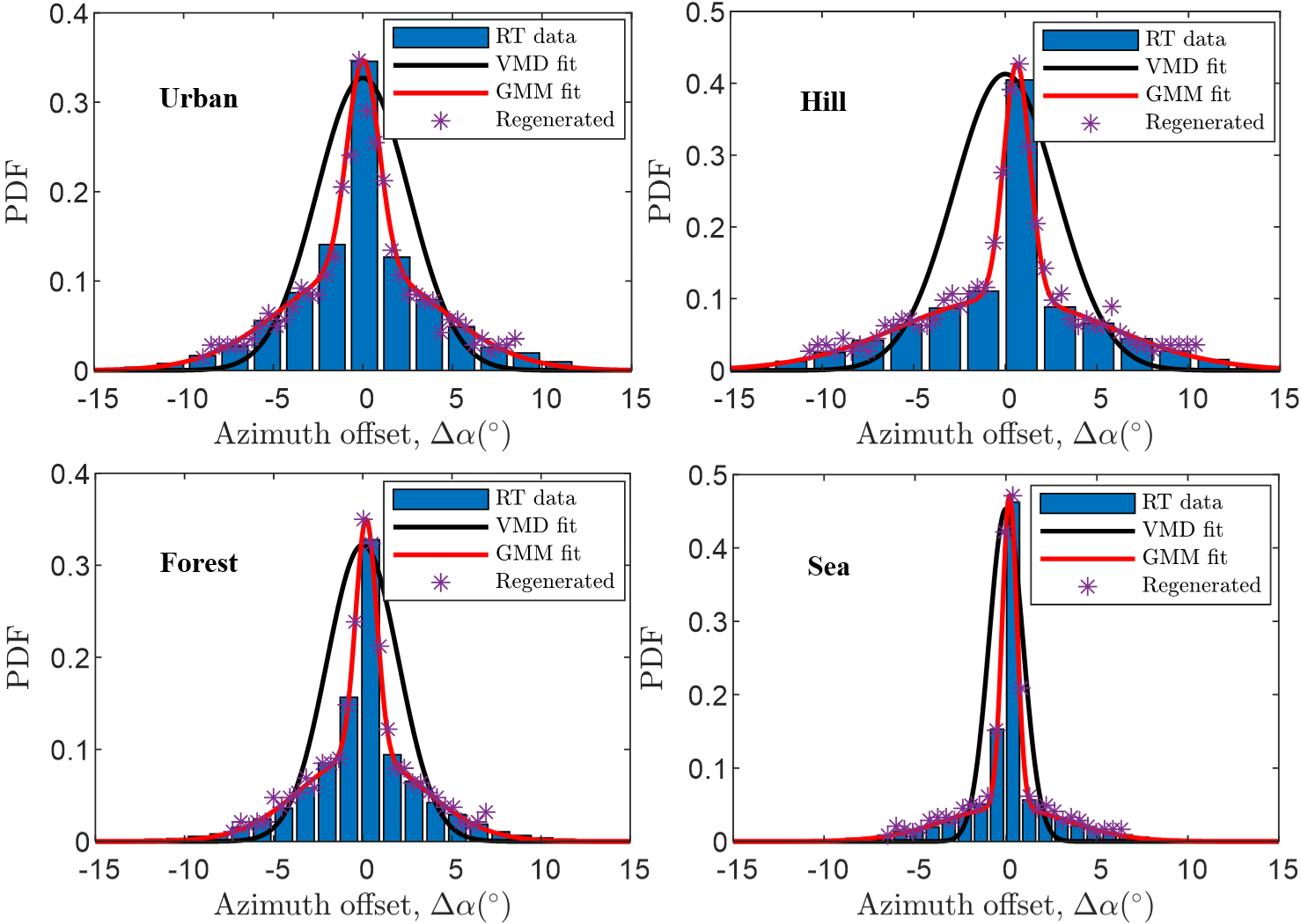}
	\caption{The PDFs of azimuth offset.}
    \label{fig:4}
\end{figure}
\begin{figure}[!t]
\setlength{\belowcaptionskip}{-0.8cm}
	\centering
	\includegraphics[width=85mm]{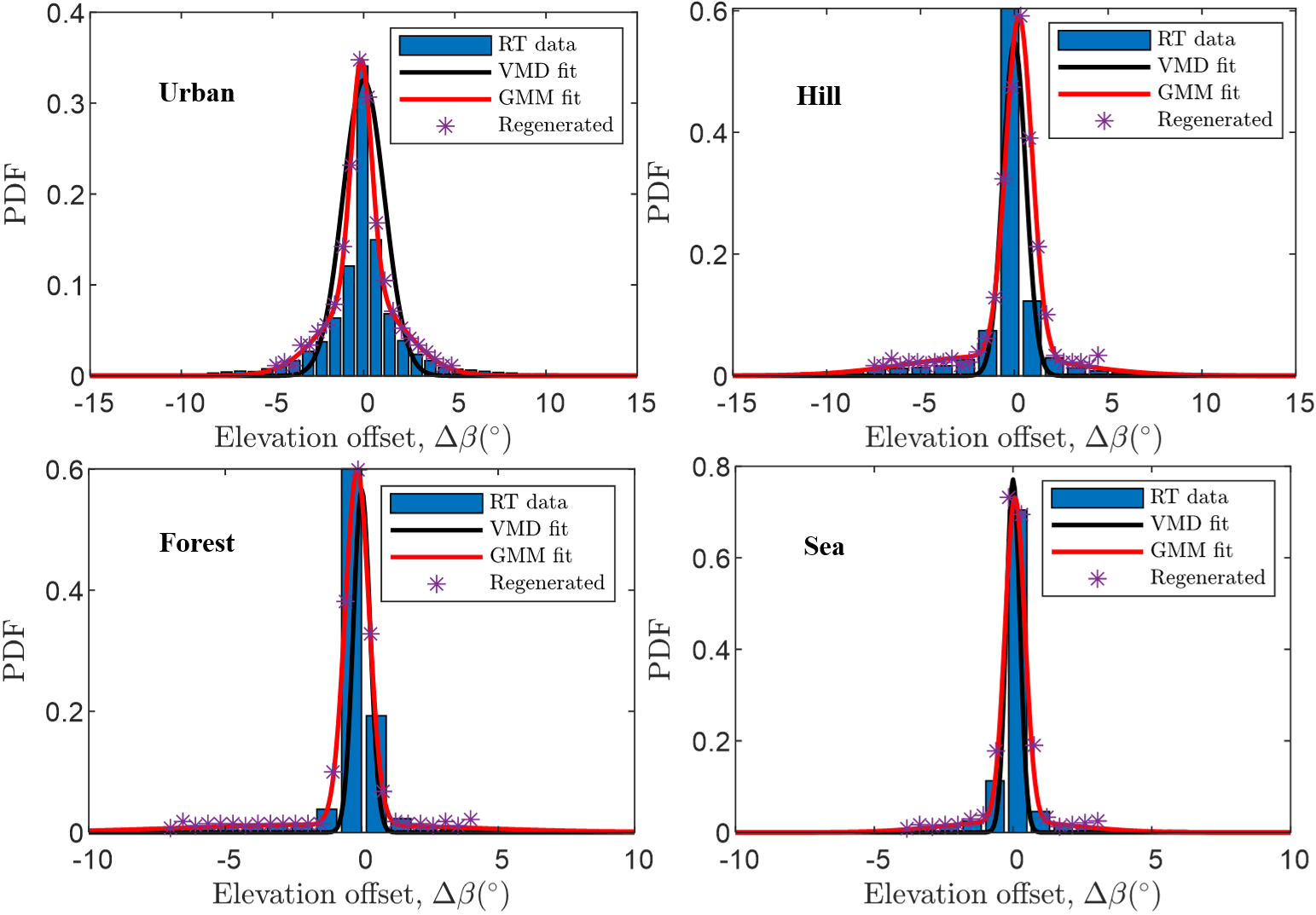}
	\caption{The PDFs of elevation offset.}
    \label{fig:5}
\end{figure}
\par For providing random ray parameters for each channel generation, we adopt a MMEA method to generate the random variables based on the above GMM distributions. Taking the ray delay offset as an example, the MMEA can be expressed as
\begin{equation}
\int_{{\tau _{k - 1}}}^{{\tau _k}} {{f_{\Delta \tau }}(\tau )} d\tau  = \frac{1}{K},{\rm{  }}k = 1,2, \ldots ,K
\label{26}
\end{equation}
\noindent where the initial value of ${\tau _0}$ is determined by the distribution range of GMM and $K$ is the total number of required parameters. Note that there is no closed-form expression of GMM integration. In this paper, the integral interval is divided into I small parts and the equation (27) can be equivalently rewritten as
\begin{equation}
\sum\limits_{i = 0}^{I - 1} {f\left( {\frac{{i{\tau _k} + \left( {I - i} \right){\tau _{k - 1}}}}{I}} \right)} \frac{{{\tau _k} - {\tau _{k - 1}}}}{I} = \frac{1}{K}
\label{27}
\end{equation}
\par To verify the effectiveness of MMEA method, we perform this method to generate 100000 samples of each kind of parameters under four scenarios. The statistical distributions of generated data are plotted with the fitting curve in Fig. 3 $-$ Fig. 5 for comparison purpose. As we can see that the PDFs of regenerated data are consistent with previous obtained GMM fitting curves under four scenarios. In previous work, we have found that the power in dB can be expressed as a linear expression with respect to the delay within each path \cite{Mao20_Sensors}. So, the ray power offset can be calculated by
\begin{equation}
\Delta {P_{n,m}} = {\log _{10}}({\rm{e}})\zeta \Delta {\tau _{n,m}}
\label{28}
\end{equation}
\noindent where $\zeta $ is a linear slope that depends on the scenarios. Finally, the channel parameter of ray power can be obtained by
\begin{equation}
{P_{n,m}}(t) = {\bar P_n}(t) + \Delta {P_{n,m}}.
\label{29}
\end{equation}

\section{Simulation results and analyze}
To verify the correctness of proposed channel model, the UAV mmWave channels under a typical urban scenario are simulated at 28 GHz. The satellite view and reconstructed digital map of simulation scenario are shown in Fig. 4. The UAV flies at the constant height of 150 m with the velocity of 10 m/s. In the urban scenario, the buildings with average height of 30 m are densely distributed and the main material of buildings are concrete. Moreover, the soil moisture of the scenarios is considered as well.Theproposed U2V mmWave channel model at 28 GHz under the urban scenario is generated and validated in this section. The satellite view and reconstructed scenario for channel generation are shown in Fig. 6. The UAV flies at the constant height of 150 m with the velocity of 10 m/s and the ground Rx moves with the velocity of 1 m/s. Both of the UAV and Rx are equipped with omnidirectional antennas. For the undergoing urban scenario, the buildings with average height of 30 m are densely distributed and the main material of bui ldings are assumed to be concrete.
\begin{figure*}[!t]
\setlength{\belowcaptionskip}{-0.5cm}
\centering
\subfigure[] {\includegraphics[width=55mm]{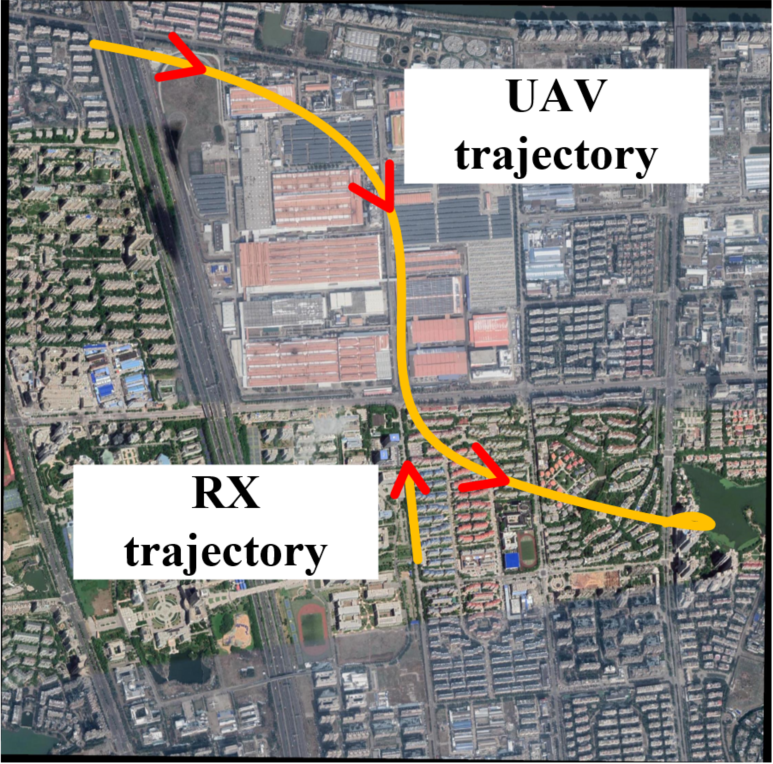}}
\hspace{.20in}
\subfigure[] {\includegraphics[width=65mm]{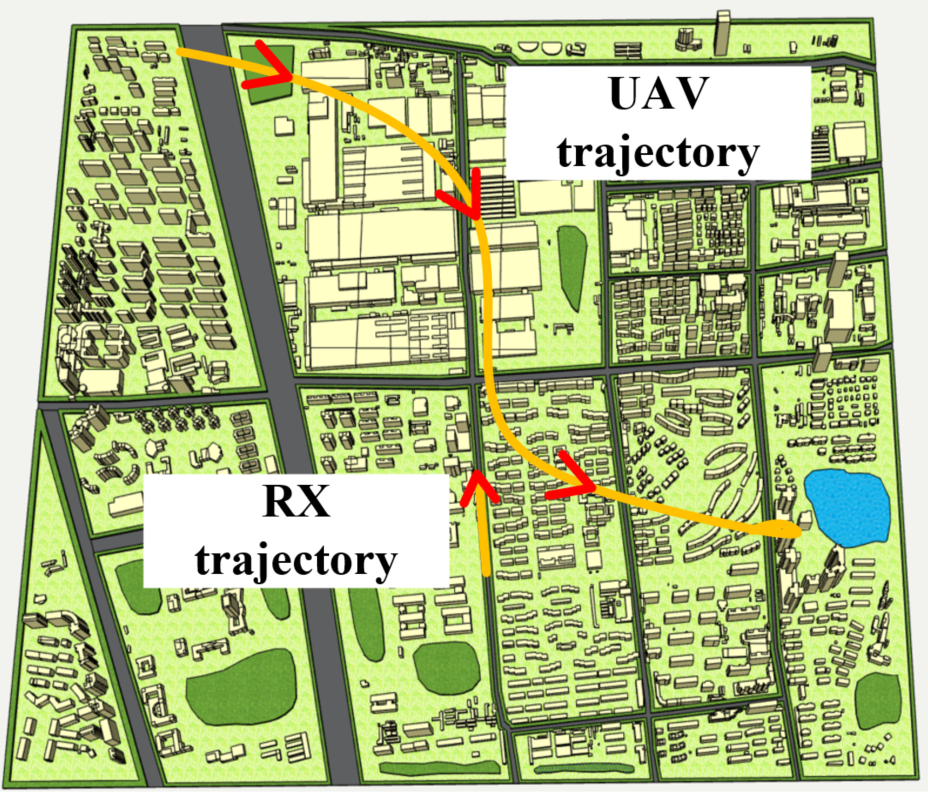}}
\caption{ (a) Satellite view and (b) reconstructed urban scenario.  }
\label{fig:6}
\end{figure*}
\par The time-variant 3D PDPs of generated channel along the flight trajectory is given in Fig. 7. The absolute time delay is used to directly show the delay varying along with the movement of UAV. In Fig. 7, we can find that when the distance between UAV and ground receiver is shorter, the path delay is smaller and the power is bigger, and vice versa. Moreover, it can be seen that the LoS path only exists between about 80 s and 180 s due to the obstacle of high buildings. There is an NLoS path closely near the LoS path which is the ground reflected path. Moreover, some other NLoS paths appear or disappear in different time instant, which also demonstrate the birth-and-death process of propagation paths in UAV communications \cite{Zhu18_CC}.
\vspace{0.2cm}
\setlength{\belowcaptionskip}{-0.3cm}
\setlength{\abovecaptionskip}{0.1cm}
\begin{figure}[!b]
	\centering
	\includegraphics[width=85mm]{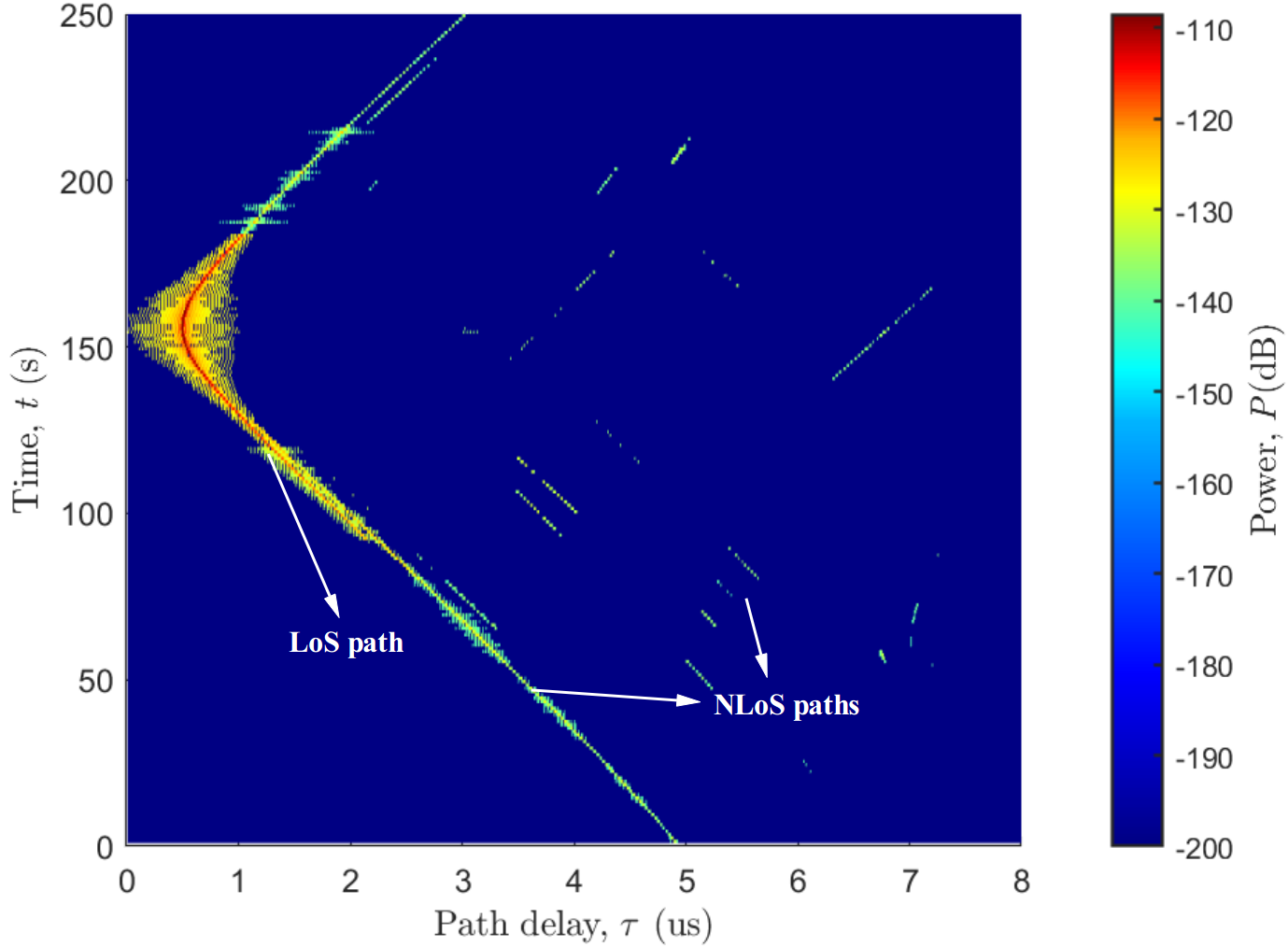}
	\caption{The generated time-variant PDPs. }
    \label{fig:7}
\end{figure}
\setlength{\abovedisplayskip}{0.5cm}
\par The second order statistical properties, i.e., ACF and DPSD, are often used to evaluate the effectiveness of channel models. The normalized ACF can be expressed as
\begin{equation}
r\left( {t;\Delta t} \right) = \frac{{{\rm{E}}\left[ {{h^*}(t)h(t + \Delta t)} \right]}}{{\sqrt {{\rm{E}}\left[ {{{\left| {{h^*}(t)} \right|}^2}} \right]{\rm{E}}\left[ {{{\left| {h(t + \Delta t)} \right|}^2}} \right]} }}
\label{30}
\end{equation}
\noindent where ${\rm{E}}\left(  \cdot  \right)$ is the expectation function and ${\left(  \cdot  \right)^ * }$ is the complex conjugate. The DPSD can be obtained by the Fourier transform on ACF. The time-variant DPSD of proposed channel generation method is given in Fig. 8(a).  It can be seen that when the UAV flies over the receiver, the Doppler frequency of LoS path rapidly turns into negative after 150 s. The Doppler frequencies of NLoS paths are complicated due to the 3D scattering environment and 3D trajectory of UAV communication scenario. For comparison purpose, based on the full RT method on the reconstructed urban scenario, the calculated DPSD is also shown in Fig. 8(b). Due to randomness of parameter generation, the calculated result can only be qualitatively compared with the generated one. It shows that the DPSD of proposed method has the same trend with the ones of RT simulation. To further verify the consistency of proposed method with the realistic channel, the ACF is simulated in the similar scenario as \cite{GuanK20_Access}  at 24 GHz, and the comparison results in Fig. 9 show that the ACF of proposed channel generator is consistent with the measured one.

\vspace{-0.5cm}  
\setlength{\belowcaptionskip}{0.3cm}   
\begin{figure}[htbp]
\setlength{\belowcaptionskip}{-0.5cm}
\centering
\subfigure[] {\includegraphics[width=85mm,height=66mm]{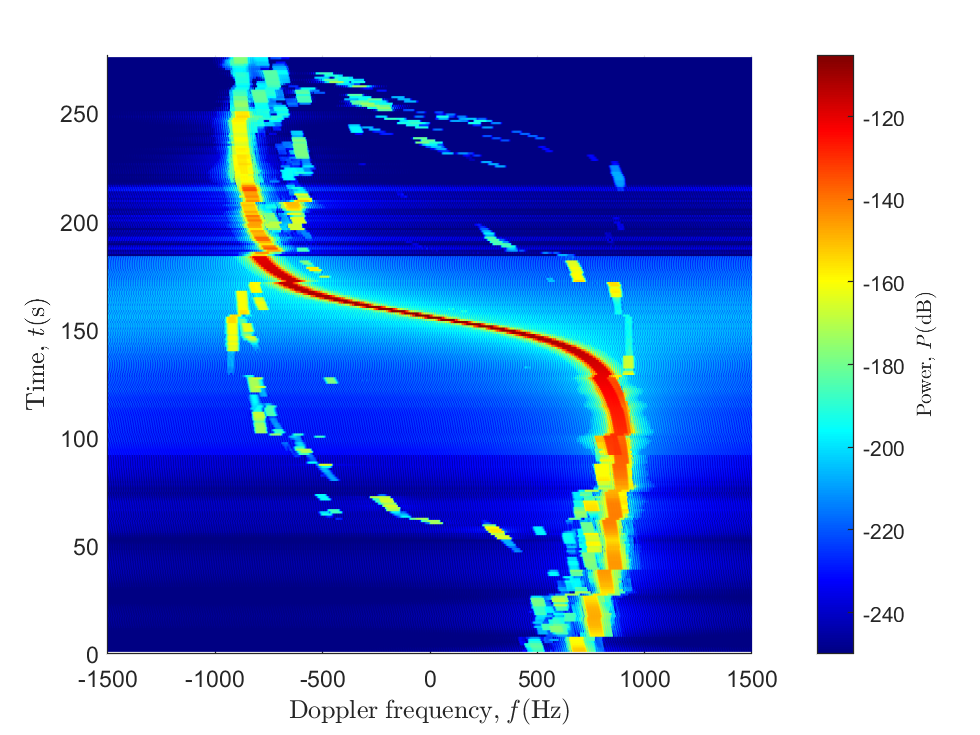}}
\subfigure[] {\includegraphics[width=85mm,height=66mm]{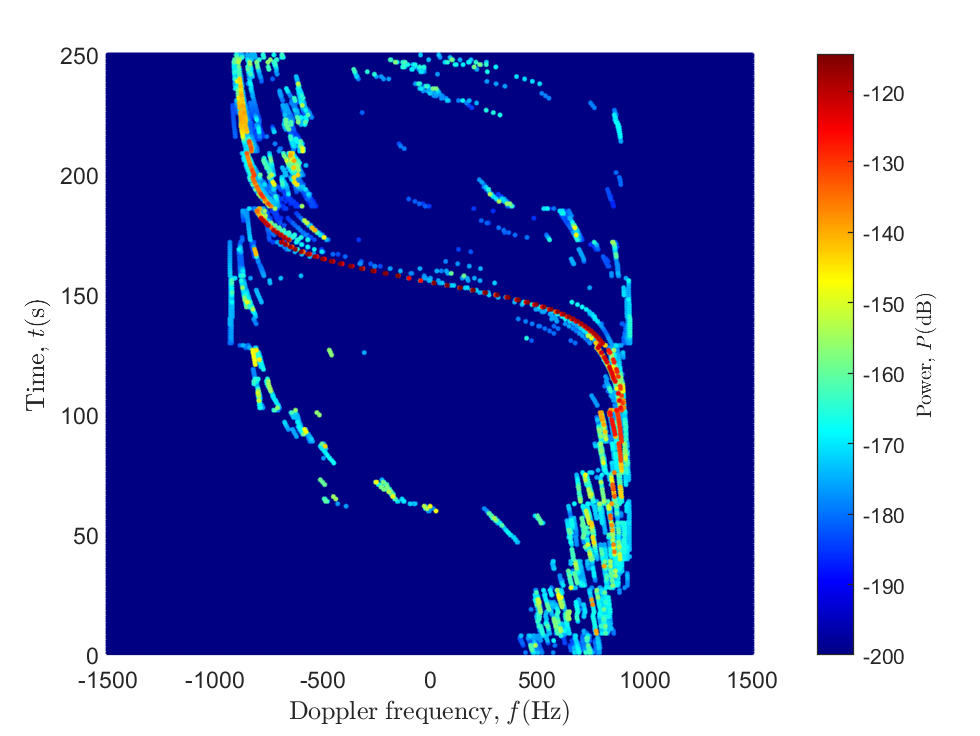}}
\caption{The time-variant DPSDs of (a) proposed channel generator and (b) deterministic RT method.}
\label{fig:8}
\end{figure}
\begin{figure}[htbp]
\setlength{\belowcaptionskip}{-0.5cm}
	\centering
	\includegraphics[width=85mm]{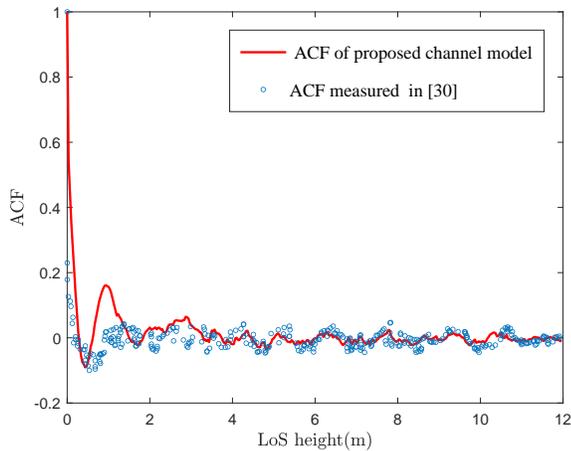}
	\caption{The generated and measured ACFs.}
    \label{fig:9}
\end{figure}

\section{Conclusions}
This paper has proposed a non-stationary 3D U2V mmWave channel model considering 3D scattering, 3D trajectory, and 3D antenna array. Meanwhile, a map-based parameter computation and generation method has been developed, which could guarantee both the efficiency and precision. The inter-path parameters have been accurately calculated in a deterministic way based on the geometric parameters. The intra-path parameters have been generated by the fitting GMM according to the massive channel data. Moreover, a simplified map reconstruction method has also given to reduce the complexity of channel generation. At last, the statistical properties of proposed channel model, i.e., PDP, ACF, and DPSD, have been generated and verified with the calculated and measured results. In the future, we will perform channel measurements to optimize the intra-path parameters as well as the channel model.
\ifCLASSOPTIONcaptionsoff
  \newpage
\fi

\end{document}